\documentclass{article}

\usepackage[british]{babel}
\usepackage[T1]{fontenc}
\usepackage[ansinew]{inputenc}
\usepackage[all]{xypic}
\usepackage{amsmath,amssymb}
\usepackage{a4}
\usepackage{Common/prooftree}
\usepackage{stmaryrd}
\usepackage{graphicx}
\usepackage{latexsym}
\usepackage{theorem}

\long\def\ignore#1{\relax}

\newcommand\struto{\raise6pt\hbox{\strut}}%
\newcommand\strutb{\raise-6pt\hbox{\strut}}%

\newtheorem{theorem}{Theorem}
\newtheorem{lemma}[theorem]{Lemma}
\newtheorem{corollary}[theorem]{Corollary}

{\theorembodyfont{\upshape}      
  \newtheorem{definition}{Definition}}
{\theorembodyfont{\upshape}      
  }
{\theorembodyfont{\upshape}      
  }
{\theorembodyfont{\upshape}      
  \newtheorem{remark}[theorem]{Remark}}

\newtheorem{proposition}{Proposition}

\newcommand\pushright[1]{{              
    \parfillskip=0pt            
    \widowpenalty=1000000000         
    \displaywidowpenalty=10000  
    \finalhyphendemerits=0      
    \leavevmode                 
    \unskip                     
    \nobreak                    
    \hfil                       
    \penalty50                  
    \hskip.2em                  
    \null                       
    \hfill                      
    {#1}                        
    \par}}                      

\newcommand{\qedsymbol}{$\Box$}

\newenvironment{proof}{\noindent{\bf Proof: }}
{\pushright{\qedsymbol}\penalty-700 \smallskip}

\renewcommand{\paragraph}[1]{\mbox{}\\{\bf \large #1}}

\newenvironment{ann}[3][{}]
{\begin{trivlist}\item[]
    {\bf {{#2}~\ref{#3}\ #1}}} 
{
\end{trivlist}}

\newcommand{\eqdef}{:=\ }

\renewcommand\b{\beta}

\renewcommand\l{\lambda}

\newcommand\lC{$\l_{\textsf C}$}

\newcommand\mathFomega{F_\omega}
\newcommand\Fomega{\ifmmode\mathFomega\else$\mathFomega$\fi}
\newcommand\mathFomegaC{F_\omega^{\mathcal C}}
\newcommand\FomegaC{\ifmmode\mathFomegaC\else$\mathFomegaC$\fi}
\newcommand\mathDNE{\mathrm{DNE}}
\newcommand\DNE{\ifmmode\mathDNE\else$\mathDNE$\fi}

\newcommand{\ie}{i.e.\ }

\newcommand\red{\rightarrow}

\newcommand\inve[1]{{#1}^{-1}}
\newcommand\co[2]{{#1}\cdot{#2}}
\newcommand\Id[1][]{\textsf{Id}_{#1}}
\newcommand\suc[2]{\hspace{0.1cm}#1\hspace{-0.1cm}(#2)}

\newcommand\patriarchal{patriarchal}

\newcommand\rf[1]{\textsf{rf}^{#1}}
\newcommand\nf[1]{\textsf{nf}^{#1}}
\newcommand\SN[1]{\textsf{SN}^{#1}}
\newcommand\BN[2]{\textsf{BN}^{#1}_{#2}}

\ignore{

}

\newcommand{\sep}{\mbox{$\;|\;$}}    

\newcommand\subtermeq{\sqsubseteq}

\newcommand\FV[2][{}]{\textsf{FV}_{#1}(#2)}

\newcommand{\subst}[3]{ \left\{{}^{#3}\hspace{-6pt}\diagup\hspace{-2pt}_{#2} \right\}\hspace{-1pt} #1 }

\newcommand{\Rew}[2][]{\stackrel{#1}{\longrightarrow}_{#2}\;}
\newcommand{\Rewn}[2][*]{{\longrightarrow}^{#1}_{#2}\;}
\newcommand{\Rewplus}[2][+]{{\longrightarrow}^{#1}_{#2}\;}

\newcommand\daggerL{\raise3pt\hbox{\rotatebox{-40}{$\dagger$}}}
\newcommand\daggerR{\raise0pt\hbox{\rotatebox{40}{$\dagger$}}}

\ignore{

}

\newcommand\sigm{\textsf{assoc}}
\newcommand\sigmm{\overline{\textsf{assoc}}}
\newcommand\activ{\textsf{act}}

\newcommand\ml[3]{(\overline\lambda #1.#2)\ #3}
\newcommand\mlet[3]{\textsf{let } #1=#3\textsf{ in }#2}
\newcommand\mlname{\overline\lambda}
\newcommand\beti{\beta1}
\newcommand\betii{\beta2}
\newcommand\betiii{\beta12}
\newcommand\betbl{\overline\beta}
\newcommand\betbk{\beta\kappa}
\newcommand\betb{\beta\overline\kappa}
\newcommand\gred{\hookrightarrow}
\newcommand\knco{\leadsto_1}
\newcommand\kaco{\leadsto_2}

\newcommand\fgtname{\phi}
\newcommand\fgt[1]{\fgtname(#1)}
\newcommand\REW[1]{\rightharpoonup_{#1}}

\newcommand\full[2][{}]{\ignore{#2}{#1}}

\begin{document}

\title{Termination of $\l$-calculus\\ with an extra call-by-value rule}
\author{Stéphane Lengrand${}^{1,2}$\\
${}^{1}$CNRS, Ecole Polytechnique, France\\
${}^{2}$University of St Andrews, Scotland\\
\texttt{Lengrand@LIX.Polytechnique.fr}
}
\date{26th November 2007}

\maketitle

Notations and standard results are presented in
Appendix~\ref{section:reminder}.

We consider the following rule in $\l$-calculus: 
$$\sigm\quad {(\l x.M)\ ((\l y.N)\ P)}\Rew{}{(\l y.(\l x.M)\ N)\ P)}$$

We want to prove
\begin{proposition}\label{prop:goal}
$\SN{\beta}{}\subseteq\SN{{\sigm\beta}}$.
\end{proposition}

\begin{lemma}\label{lem:sigmterm}
$\Rew{\sigm}$ is terminating in $\l$-calculus. 
\end{lemma}
\begin{proof}
Each application of the rule decreases by one the number of pairs of $\l$
that are not nested. 
\end{proof}

To prove Proposition~\ref{prop:goal} above, it would thus be sufficient to prove
that $\Rew{\sigm}$ could be adjourned with respect to $\Rew{\b}$, in
other words that $\co{\Rew{\sigm}}{\Rew{\b}}\subseteq
\co{\Rew{\b}}{\Rewn{\sigm\b}}$ (the adjournment technique leads
directly to the desired strong normalisation result).
When trying to prove the property by induction and case analysis on the
$\b$-reduction following the $\sigm$-reduction to be adjourned, all
cases allow the adjournment but one, namely:
$${(\l x.M)\ ((\l y.N)\ P)}\Rew{\sigm}{(\l y.(\l x.M)\ N)\ P}
\Rew{\b}{(\l y.\subst M x N)\ P}$$

Hence, we shall assume without loss of generality that the
$\beta$-reduction is not of the above kind. For that we need to
identify a sub-relation of $\b$-reduction $\gred$ such that
\begin{itemize}
\item
$\Rew{\sigm}$ can now be adjourned with respect to $\gred$
\item
we can justify that there is no loss of generality.
\end{itemize}
For this we give ourselves the possibility of marking $\l$-redexes and
forbid reductions under their (marked) bindings, so that, if in the
$\sigm$-reduction above we make sure that $(\l y.(\l x.M)\ N)\ P)$ is
marked, the problematic $\b$-reduction is forbidden.

Hence we use the usual notation for a marked redex $(\overline \l y.Q)\
P$, but we can also see it as the construct $\mlet y Q P$ of
\lC~\cite{Moggi:edinburgh1988} and other works on call-by-value
$\l$-calculus.
We start with a reminder about marked redexes.

\begin{definition}
The syntax of the $\l$-calculus is extended as follows:
$$
M,N::= x\sep \l x.M\sep M\ N\sep \ml x M N
$$
Reduction is given by the following system $\betiii$:
$$
\begin{array}{lll}
\beti&(\l x.M)\ N&\Rew{}\subst N x M\\
\betii&\ml x M N&\Rew{}\subst N x M
\end{array}
$$

The forgetful projection onto $\l$-calculus is straightforward:
$$
\begin{array}{lll}
\fgt{x}&\eqdef x\\
\fgt{\l x.M}&\eqdef \l x.\fgt{M}\\
\fgt{M\ N}&\eqdef \fgt M\ \fgt N\\
\fgt{\ml x M N}&\eqdef (\l x.\fgt M)\ \fgt N
\end{array}
$$
\end{definition}

\begin{remark}\label{rem:fgtss}
Clearly, $\Rew{\betiii}$ strongly simulates $\Rew{\beta}$ through $\fgtname^{-1}$ and 
$\Rew{\beta}$ strongly simulates $\Rew{\betiii}$ through $\fgtname$.
\end{remark}

\subsection*{Reducing under $\mlname$ and erasing $\mlname$ can be strongly adjourned}
In this section we identify the reduction notion $\gred\
(\subseteq\Rew{\betiii})$ and we argue against the loss of generality
by proving that
\full[$\co{\Rew{\betiii}}{\gred}\subseteq\co{\gred}{(\Rew{\betiii}\cup\gred)^+}$,
a strong case of adjournment, presented in
Appendix~\ref{section:newadjourn}, whose direct corollary is that, for
every sequence of $\betiii$-reduction, there is also a sequence of
$\gred$-reduction of the same length and starting from the same
term.]{$\Rew{\betiii}$ can be strongly adjourned with respect to
$\gred$.}

We thus split the reduction system $\betiii$ into two cases depending on
whether or not a reduction throws away an argument that contains some
markings:
\begin{definition}
$$
\begin{array}{cc}
\betbk&\left\{
\begin{array}{ll@{\quad}l}
 (\l x.M)\ P &\Rew{} M&\mbox{if $x\not\in\FV M$ and there is a term
   $\ml x N
  Q\subtermeq P$}\\
 \ml x M P &\Rew{} M&\mbox{if $x\not\in\FV M$ and there is a term
   $\ml x N
  Q\subtermeq P$}
\end{array}\right.\\
\betb&\left\{
\begin{array}{ll@{\quad}l}
 (\l x.M)\ P &\Rew{} M&\mbox{if $x\in\FV M$ or there is no term $\ml x N
  Q\subtermeq P$}\\
 \ml x M P &\Rew{} M&\mbox{if $x\in\FV M$ or there is no term $\ml x N
  Q\subtermeq P$}
\end{array}\right.
\end{array}
$$
\end{definition}

\begin{remark}
Clearly, $\Rew{\betiii}=\Rew{\betbk}\cup\Rew{\betb}$.
\end{remark}

No we distinguish whether or not a reduction occurs underneath a
marked redex, via the following rule and the following notion of
contextual closure:
\begin{definition}
$$
\begin{array}{lll@{\quad}l}
\betbl& \ml x M P &\Rew{} \ml x N P&\mbox{if $M \Rew{\betiii} N$}
\end{array}
$$
Now we define a weak notion of contextual closure for a rewriting
system $i$:
$$
\begin{array}c
\infer{M\REW{i} N}{i:\ M\Rew{} N}\quad
\infer{\l x.M\REW{i} \l x. N}{M\REW{i} N}\quad
\infer{M\ P\REW{i} N\ P}{M\REW{i} N}\quad
\infer{P\ M\REW{i} P\ N}{M\REW{i} N}\\\\
\infer{\ml x P M\REW{i} \ml x P N}{M\REW{i} N}
\end{array}
$$
\end{definition}

Finally we use the following abbreviations:
\begin{definition}
Let $\gred\eqdef\REW{\betb}$ and $\knco\eqdef\REW{\betbk}$ and
$\kaco\eqdef\REW{\betbl}$.
\end{definition}

\begin{remark}\label{rem:betunion}
Clearly, $\Rew{\betiii}=\gred\cup\knco\cup\kaco$.
\end{remark}

\begin{lemma}\label{lem:highestmarker}
If $\ml x N Q\subtermeq P$, then there is $P'$ such that
$P\gred P'$.
\end{lemma}
\begin{proof}
By induction on $P$
\begin{itemize}
\item The case $P=y$ is vacuous. 
\item For $P=\l y.M$, we have $\ml x N Q\subtermeq M$ and the
induction hypothesis provides $M\gred M'$, so $\l y.M\gred \l
y.M'$.
\item For $P=M_1\ M_2$, we have either $\ml x N Q\subtermeq M_1$ or
$\ml x N Q\subtermeq M_2$. In the former case the
induction hypothesis provides $M_1\gred M'_1$, so $M_1\
M_2\gred M'_1\ M_2$. The latter case is similar.
\item Suppose $P=\ml y {M_1} {M_2}$.
If there is a term $\ml {x'}{N'} {Q'}\subtermeq M_2$, the induction
hypothesis provides $M_2\gred M'_2$, so $\ml y {M_1}
{M_2}\gred\ml y {M_1} {M'_2}$.
If there is no such term $\ml {x'}{N'}{Q'}\subtermeq M_2$, we have
$\ml y {M_1} {M_2}\gred\subst{M_1} y{M_2}$.
\end{itemize}
\end{proof}

\begin{lemma}
$\knco\subseteq \co\gred{\knco}$
\end{lemma}
\begin{proof}
By induction on the reduction step $\knco$.

For the base cases $(\l x.M)\ P \Rew{\betbk} M$ or $\ml x M P
\Rew{\betbk} M$ with $x\not\in\FV M$
and $\ml y N Q\subtermeq P$, Lemma~\ref{lem:highestmarker} provides
the reduction $P\gred P'$, so $(\l x.M)\ P \gred (\l x.M)\ P' \knco M$
and $\ml x M P \gred \ml x M {P'} \knco M$.

The induction step is straightforward as the same contextual closure
is used on both sides (namely, the weak one).
\end{proof}

\begin{lemma}\label{lem:kaco}
$\co{\kaco}\gred\subseteq \co\gred{\Rewplus{\betiii}}$
\end{lemma}
\begin{proof}
By induction on the reduction step $\gred$.
\full[See appendix~\ref{section:proofs}.]{\begin{itemize}
\item For the base case where the $\betb$-reduction is a
$\betii$-reduction, we have\linebreak $M\kaco\ml x N P\gred \subst {N} x
P$ with $x\in\FV N$ or $P$
has no marked redex as a subterm. We do a case analysis on the reduction step $M\kaco\ml x N P$.

If $M=\ml x {N'} P\kaco\ml x N P$ because $N'\Rew{\betiii}N$ then
$\ml x {N'} P\gred \subst {N'} x P\Rew{\betiii}\subst {N} x P$.

If $M=\ml x N {P'}\kaco\ml x N P$ because
$P'\kaco P$, then it means that $P$ has a marked redex as a subterm, so we must
have $x\in\FV N$. Hence $\ml x N {P'}\gred\subst {N} x
{P'}\Rewplus{\betiii}\subst {N} x P$.
\item
For the base case where the $\betb$-reduction is a
$\beti$-reduction, we have\linebreak
$M\kaco(\l x.N)\ P\gred \subst {N} x P$ with $x\in\FV N$ or $P$
has no marked redex as a subterm. We do a case analysis
on the reduction step $M\kaco(\l x.N)\ P$.

If $M=M'\ P\kaco(\l x.N)\ P$ because $M'\kaco\l x.N$
then $M'$ must be of the form $\l x.M''$ with $M''\kaco N$.
Then $(\l x.M'')\ P\gred\subst {M''} x P$ (in case $P$ has a marked
subterm, notice that $x\in\FV{N}\subseteq\FV {M''}$), and $\subst
{M''} x P\Rew{\betiii}\subst {N} x P$.

If $M=(\l x.N)\ P'\kaco(\l x.N)\ P$ because $P'\kaco P$,
then it means that $P$ has a marked redex as a subterm, so we must
have $x\in\FV N$. Hence $(\l x.N)\ P'\gred\subst {N} x
{P'}\Rewplus{\betiii}\subst {N} x P$.
\item
The closure under $\lambda$ is straightforward.
\item For the closure under application, left-hand side, we have
$M\kaco N\ P\gred {N'}\ P$ with $N\gred {N'}$. We do a case analysis
on the reduction step $M\kaco N\ P$.

If $M=M'\ P \kaco N\ P$ with $M' \kaco N$, the induction
hypothesis gives $M'\co{\gred}{\Rewplus{\betiii}}N'$ and the weak contextual
closure gives $M'\ P\co{\gred}{\Rewplus{\betiii}}N'\ P$.

If $M=N\ P'\kaco N\ P$ with $P' \kaco P$, we can also
derive $N\ P'\gred {N'}\ P'\Rew{\betiii}N'\ P$.
\item For the closure under application, right-hand side, we have
$M\kaco N\ P\gred N\ P'$ with $P\gred P'$. We do a case analysis
on the reduction step $M\kaco N\ P$.

If $M=M'\ P \kaco N\ P$ with $M' \kaco N$, we can also
derive $M'\ P\gred M'\ P'\Rew{\betiii}N\ P'$.

If $M=N\ M'\kaco N\ P$ with $M' \kaco P$, the induction
hypothesis gives $M'\co{\gred}{\Rewplus{\betiii}}P'$ and the weak contextual
closure gives $N\ M'\co{\gred}{\Rewplus{\betiii}}N\ P'$.
\item For the closure under marked redex we have $M\kaco \ml
x P N \gred \ml x P {N'}$ with $N \gred {N'}$. We do a case analysis on
the reduction step $M\kaco\ml x P N $.

If $M=\ml x {P'} N \kaco \ml x P N$ because $P'
\Rew{\betiii} P$, we can also derive $\ml x {P'} N \gred \ml x {P'}
{N'}\Rew{\betiii} \ml x P {N'}$.

If $M=\ml x P {M'}\kaco \ml x P N$ with $M' \kaco
N$, the induction hypothesis gives $M'\gred Q\Rewplus{\betiii}N'$ and
the weak contextual closure gives $\ml x P {M'}\gred\ml x P
Q\Rewplus{\betiii}\ml x P{N'}$.
\end{itemize}}
\end{proof}

\begin{corollary}\label{cor:betgred}
$\Rew{\betiii}$ can be strongly adjourned with respect to $\gred$.
\end{corollary}
\begin{proof}
Straightforward from the last two theorems, and Remark~\ref{rem:betunion}.
\end{proof}

\subsection*{$\sigm$-reduction}
We introduce two new rules in the marked
$\l$-calculus to simulate $\sigm$:
$$
\begin{array}{lll}
\sigmm& {\ml x M {\ml y N P}}&\Rew{}{\ml y {\ml x M N} P}\\
\activ& (\l x.M)\ N&\Rew{}\ml x M N
\end{array}
$$

\begin{remark}\label{rem:fgtss2}
Clearly, $\Rew{\sigmm\activ}$ strongly simulates $\Rew{\sigm}$ through
$\fgtname^{-1}$.
\end{remark}

Notice that with the $\mlet{}{}{}$-notation, $\sigmm$ and $\activ$ are
simply the rules of \lC
$$
\begin{array}{lll}
\sigmm& {\mlet x M {(\mlet y N P)}}&\Rew{}{\mlet y {\mlet x M N} P}\\
\activ& (\l x.M)\ N&\Rew{}\mlet x M N
\end{array}
$$

\begin{lemma}\label{lem:sigmgred}
$\co{\Rew{\sigmm\activ}}{\gred}\subseteq \co{\gred}{\Rewn{\sigmm\activ}}$
\end{lemma}
\begin{proof}
By induction on the reduction step $\gred$.
\full[See appendix~\ref{section:proofs}.]{\begin{itemize}
\item For the first base case, we have $M\Rew{\sigmm\activ}(\l x.N)\
P\gred\subst N x P$ with $x\in\FV N$ or $P$ has no marked subterm. 
Since root $\sigmm\activ$-reduction produces
neither $\l$-abstractions nor applications at the root, note that $M$
has to be of the form $(\l x.N')\ {P'}$, with either
$N'\Rew{\sigmm\activ}N$ (and $P'=P$) or $P'\Rew{\sigmm\activ}P$ (and
$N'=N$).
In both cases, $x\in\FV N\subseteq\FV {N'}$ or $P'$ has no marked
subterm, so we also have $(\l x.N')\ {P'}\gred
\subst{N'}x{P'}\Rewn{\sigmm\activ}\subst N x P$.
\item For the second base case, we have $M\Rew{\sigmm\activ}\ml x N
P\gred\subst N x P$ with $x\in\FV N$ or $P$ has no marked subterm.  We
do a case analysis on $M\Rew{\sigmm\activ}\ml x N P$.

If $M=\ml {x'}{M_1} {\ml x {M_2} P}\Rew{\sigmm}
\ml x {\ml {x'}{M_1} {M_2}} P$ with $N=\ml {x'}{M_1} {M_2}$, we also
have $M=\ml {x'}{M_1} {\ml x {M_2} P}\gred \ml {x'}{M_1} {\subst
{M_2}{x}P}=\subst N x P$.

If $M=(\l x.N)\ P\Rew{\activ}\ml x N P$ then  
$M\gred\subst N x P$.

If $M=\ml x {N'} {P'}\Rew{\sigmm\activ}\ml x N P$ with either
$N'\Rew{\sigmm\activ}N$ (and $P'=P$) or $P'\Rew{\sigmm\activ}P$ (and
$N'=N$), we have, in both cases, $x\in\FV N\subseteq\FV {N'}$ or $P'$
has no marked subterm, so we also have $(\l x.N')\ {P'}\gred
\subst{N'}x{P'}\Rewn{\sigmm\activ}\subst N x P$.
\item
The closure under $\l$ is straightforward.
\item
For the closure under application, left-hand side, we have
$Q\Rew{\sigmm\activ}M\ N\gred M'\ N$ with $M\gred M'$. We
do a case analysis on
$Q\Rew{\sigmm\activ}M\ N$.

If $Q=M''\ N\Rew{\sigmm\activ}M\ N$ with $M''\Rew{\sigmm\activ}M$, the
induction hypothesis provides $M''\co{\gred}{\Rewn{\sigmm\activ}}M'$
so $M''\ N\co{\gred}{\Rewn{\sigmm\activ}}M'\ N$.

If $Q=M\ {N'}\Rew{\sigmm\activ}M\ N$ with $N'\Rew{\sigmm\activ}N$, we
also have $M\ {N'}\gred M'\ {N'}\Rew{\sigmm\activ}M'\ N$.
\item
For the closure under application, right-hand side, we have
$Q\Rew{\sigmm\activ}M\ N\gred M\ {N'}$ with $N\gred {N'}$. We
do a case analysis on
$Q\Rew{\sigmm\activ}M\ N$.

If $Q=M'\ N\Rew{\sigmm\activ}M\ N$ with $M'\Rew{\sigmm\activ}M$, we
also have $M'\ N\gred M'\ {N'}\Rew{\sigmm\activ}M\ {N'}$.

If $Q=M\ N''\Rew{\sigmm\activ}M\ N$ with $N''\Rew{\sigmm\activ}N$, the
induction hypothesis provides $N''\co{\gred}{\Rewn{\sigmm\activ}}N'$
so $M\ N''\co{\gred}{\Rewn{\sigmm\activ}}M\ N'$.
\item For the closure under marked redex, the $\gred$-reduction can
only come from the right-hand side because of the weak contextual
closure ($\gred$ does not reduce under $\mlname$), so we have
$Q\Rew{\sigmm\activ}\ml y M P\gred\ml y M {P'}$ with $P\gred {P'}$.
We do a case analysis on $Q\Rew{\sigmm\activ}\ml y M P$.

If $Q={\ml x {M'} {\ml y N P}}\Rew{\sigmm}{\ml y {\ml x {M'} N}
P}$ with $M=\ml x {M'} N$, we also have
$Q={\ml x {M'} {\ml y N P}}\gred{\ml x {M'} {\ml y N {P'}}}
\Rew{\sigmm}{\ml y {\ml x {M'} N} {P'}}$.

If $Q=(\l y.M)\ P\Rew{\activ}\ml y M P$, then we also have 
$Q=(\l y.M)\ P\gred (\l y.M)\ {P'}\Rew{\activ}\ml y M {P'}$.
\end{itemize}}
\end{proof}

\begin{lemma}
$\co{\Rewn{\sigmm,\activ}}{\Rew{\betiii}}$ can be strongly adjourned with respect
to ${\gred}$.
\end{lemma}
\begin{proof}
We prove that $\forall k,\co{\co{\Rewn[k]{\sigmm,\activ}}{\Rew{\betiii}}}{\gred}\subseteq
\co{\gred}{\co{\Rewn{\sigmm,\activ}}{\Rew{\betiii}}}$ by induction on
$k$.
\begin{itemize}
\item
For $k=0$, this is Corollary~\ref{cor:betgred}.
\item
Suppose it is true for $k$. By the induction hypthesis we get
$$\co{\Rew{\sigmm,\activ}}{\co{\co{\Rewn[k]{\sigmm,\activ}}{\Rew{\betiii}}}{\gred}}\subseteq\co{\Rew{\sigmm,\activ}}{\co{\gred}{\co{\Rewn{\sigmm,\activ}}{\Rew{\betiii}}}}$$
Then by Lemma~\ref{lem:sigmgred} we get
$$\co{\Rew{\sigmm,\activ}}{\co{\gred}{\co{\Rewn{\sigmm,\activ}}{\Rew{\betiii}}}}\subseteq
\co{\co{\gred}{\Rew{\sigmm,\activ}}}{\co{\Rewn{\sigmm,\activ}}{\Rew{\betiii}}}$$
\end{itemize}
\end{proof}

\begin{remark}
Note from Lemma~\ref{lem:highestmarker} that
$\nf{\gred}\subseteq\nf{\knco\cup\kaco}\subseteq\nf{\Rew{\betiii}}\subseteq\nf{\co{\Rewn{\sigmm,\activ}}{\Rew{\betiii}}}$.
\end{remark}

\begin{theorem}
$\BN{\gred}{}\subseteq\BN{\co{\Rewn{\sigmm,\activ}}{\Rew{\betiii}}}{}$
\end{theorem}
\begin{proof}
We apply Theorem~\ref{th:adjbound}, since $\nf{\gred}\subseteq
\nf{\co{\Rewn{\sigmm,\activ}}{\Rew{\betiii}}}$ and clearly
$$(\co{\Rewn{\sigmm,\activ}}{\Rew{\betiii}})\cup{\gred}=\co{\Rewn{\sigmm,\activ}}{\Rew{\betiii}}$$
\end{proof}

\begin{theorem}
$\BN{\beta}{}\subseteq{\BN{\co{\Rewn{\sigm}}{\Rew{\beta}}}{}}$
\end{theorem}
\begin{proof}
Since $\Rew{\b}$ strongly simulates
$\gred$ through $\fgtname$, we have
$\fgtname^{-1}(\BN{\beta}{})\subseteq\BN{\gred}{}\subseteq{\BN{\co{\Rewn{\sigmm,\activ}}{\Rew{\betiii}}}{}}$.
Hence 
$\fgt{\fgtname^{-1}(\BN{\beta}{})}\subseteq\fgt{\BN{\co{\Rewn{\sigmm,\activ}}{\Rew{\betiii}}}{}}$.
Since $\fgtname$ is surjective,
$\BN{\beta}{}=\fgt{\fgtname^{-1}(\BN{\beta}{})}$.
Hence 
$\BN{\beta}{}\subseteq\fgt{\BN{\co{\Rewn{\sigmm,\activ}}{\Rew{\betiii}}}{}}$.
Also, $\co{\Rewn{\sigmm,\activ}}{\Rew{\betiii}}$ strongly simulates $\co{\Rewn{\sigm}}{\Rew{\beta}}$ through
$\fgtname^{-1}$, so\linebreak
$\fgt{\BN{\co{\Rewn{\sigmm,\activ}}{\Rew{\betiii}}}{}}\subseteq{\BN{\co{\Rewn{\sigm}}{\Rew{\beta}}}{}}$.
\end{proof}

\begin{theorem}\label{th:sbb}
$\SN{\beta}{}\subseteq\SN{{\sigm\beta}}$
\end{theorem}
\begin{proof}
First, from Lemma~\ref{lem:incl}, ${\BN{\co{\Rewn{\sigm}}{\Rew{\beta}}}{}}\subseteq{\SN{\co{\Rewn{\sigm}}{\Rew{\beta}}}{}}$.
Then from Lemma~\ref{lem:sigmterm}, $\Rew{\sigm}$ is terminating and hence
$\SN{{\sigm}}$ is stable under $\Rew{\beta}$. Hence we can apply
Lemma~\ref{lem:lexic} to get
$\SN{{\sigm\beta}}=\SN{\co{\Rewn{\sigm}}{\Rew{\beta}}}{}$.  From the
previous theorem we thus have
$\BN{\beta}{}\subseteq\SN{{\sigm\beta}}$.  Now, noticing that
$\b$-reduction in $\l$-calculus is finitely branching,
Lemma~\ref{lem:BNSN} gives $\BN{\beta}{}=\SN{\beta}{}$ and thus
$\SN{\beta}{}\subseteq\SN{{\sigm\beta}}$.
\end{proof}

\bibliographystyle{Common/good}
\bibliography{Common/abbrev-short,Common/Main,Common/crossrefs}

\appendix
\section{Reminder: Notations, Definitions and
Basic Results} 
\label{section:reminder}
\begin{definition}[Relations]$ $
\begin{itemize}
\item
We denote the composition of relations by
$\co{}{}$, the identity relation by $\Id$, and the inverse
of a relation by $\inve{}$.
\item
If $\mathcal D\subseteq\mathcal A$, we write $\mathcal R(\mathcal
D)$ for $\{M\in\mathcal B|\;\exists N\in\mathcal D,N\mathcal R
M\}$, or equivalently $\bigcup_{N\in\mathcal D}\{M\in\mathcal
B|\;N\mathcal R M\}$. When $\mathcal D$ is the singleton $\{M\}$,
we write $\mathcal R(M)$ for $\mathcal R(\{M\})$.
\item
We say that a relation $\mathcal R:\mathcal A\longrightarrow
\mathcal B$ is \emph{total} if $\mathcal R^{-1}(\mathcal
B)=\mathcal A$.
\end{itemize}
\end{definition}

\begin{remark}
Composition is associative, and identity relations are
neutral for the composition operation.
\end{remark}

\begin{definition}[Reduction relation]
$ $\begin{itemize}
\item
A \emph{reduction relation} on $\mathcal A$ is a relation from
$\mathcal A$ to $\mathcal A$.
\item Given a reduction relation $\rightarrow$ on $\mathcal A$, we
define the set of $\rightarrow$-\emph{reducible
forms} (or just \emph{reducible forms} when the relation is clear) as
 $\rf\rightarrow\eqdef\{M\in\mathcal A|\;\exists N \in\mathcal
A,M\rightarrow N\}$.  We define the set of \emph{normal forms} as $\nf\rightarrow\eqdef\{M\in\mathcal
A|\;\not\exists N \in\mathcal A,M\rightarrow N\}$.
\item
Given a reduction relation $\rightarrow$ on $\mathcal A$, we write
$\leftarrow$ for $\inve\rightarrow$, and we define
$\rightarrow^n$ by induction on the natural number $n$ as follows:\\
$\rightarrow^0\eqdef\Id$\\
$\rightarrow^{n+1}\eqdef\co{\rightarrow}{\rightarrow^n}(=\co{\rightarrow^n}
{\rightarrow})$\\
$\rightarrow^+$ denotes the transitive closure of
$\rightarrow$ (i.e. $\rightarrow^+\eqdef\bigcup_{n\geq
1}\rightarrow^n$).\\ $\rightarrow^*$ denotes the transitive and
reflexive closure of $\rightarrow$ (i.e.
$\rightarrow^*\eqdef\bigcup_{n\geq 0}\rightarrow^n$).\\ 
$\leftrightarrow$
denotes the symmetric closure of
$\rightarrow$ (i.e. $\leftrightarrow\eqdef\leftarrow\cup\rightarrow$).\\
$\leftrightarrow^*$
denotes the transitive, reflexive and symmetric closure of
$\rightarrow$.
\item
An \emph{equivalence relation} on $\mathcal A$ is a transitive,
reflexive and symmetric reduction relation on $\mathcal A$, \ie a
relation $\rightarrow\ =\ \leftrightarrow^*$, hence denoted more often by
$\sim$, $\equiv$\ldots
\item
Given a reduction relation $\rightarrow$ on $\mathcal A$ and a
subset $\mathcal B\subseteq\mathcal A$, the \emph{closure of $\mathcal B$ under $\rightarrow$} is
  $\suc{\rightarrow^*}{\mathcal B}$.
\end{itemize}
\end{definition}

\begin{definition}[Finitely branching relation]
A reduction relation $\rightarrow$ on $\mathcal A$ is
\emph{finitely branching} if $\forall M\in\mathcal A$,
$\suc\rightarrow M$ is finite.
\end{definition}

\begin{definition}[Stability]\label{def:stable}
Given a reduction relation $\rightarrow$ on $\mathcal A$, we say that
a subset $\mathcal T$ of $\mathcal A$ is
$\rightarrow$-\emph{stable} (or \emph{stable under} $\rightarrow$)
if $\suc\rightarrow{\mathcal T}\subseteq\mathcal T$.
\end{definition}

\begin{definition}[Strong simulation]\label{Def:Simul}$ $\\
Let $\mathcal R$ be a relation between two sets $\mathcal A$ and
$\mathcal B$, respectively equipped with the reduction relations
$\rightarrow_{\mathcal A}$ and $\rightarrow_{\mathcal B}$.

\emph{$\rightarrow_{\mathcal B}$ strongly simulates
$\rightarrow_{\mathcal A}$ through $\mathcal R$} if
$(\co{\inve{\mathcal R}}{\rightarrow_{\mathcal A}})\subseteq
(\co{\rightarrow_{\mathcal B}^+}{\inve{\mathcal R}})$.
\end{definition}

\begin{remark}$ $
\begin{enumerate}
\item
If $\rightarrow_{\mathcal B}$ strongly  simulates
$\rightarrow_{\mathcal A}$ through
$\mathcal R$, and if
$\rightarrow_{\mathcal B}\subseteq\rightarrow'_{\mathcal B}$ and
$\rightarrow'_{\mathcal A}\subseteq\rightarrow_{\mathcal A}$,
then $\rightarrow'_{\mathcal B}$ strongly  simulates
$\rightarrow'_{\mathcal A}$ through
$\mathcal R$.
\item If $\rightarrow_{\mathcal B}$ strongly  simulates
$\rightarrow_{\mathcal A}$ and $\rightarrow'_{\mathcal A}$ through
$\mathcal R$, then it also strongly  simulates
$\co{\rightarrow_{\mathcal A}}{\rightarrow'_{\mathcal A}}$ through $\mathcal R$.
\item
Hence, if $\rightarrow_{\mathcal B}$ strongly simulates
$\rightarrow_{\mathcal A}$ through $\mathcal R$, then it also strongly
simulates $\rightarrow^+_{\mathcal A}$ through $\mathcal R$.
\end{enumerate}
\end{remark}

\begin{definition}[Patriarchal]
Given a reduction relation $\rightarrow$ on $\mathcal A$, we say that
\begin{itemize}
\item a subset $\mathcal T$ of $\mathcal A$ is
$\rightarrow$-\emph{\patriarchal} (or just \emph{\patriarchal}
when the relation is clear) if $\forall N\in\mathcal
A,\suc\rightarrow N\subseteq\mathcal T\Rightarrow N\in\mathcal T$.
\item
a predicate $P$ on $\mathcal A$ is \emph{patriarchal}
if $\{M\in\mathcal A|\;P(M)\}$ is
\emph{patriarchal}.
\end{itemize}
\end{definition}

\begin{definition}[Normalising elements]
Given a reduction relation $\rightarrow$ on $\mathcal A$,
the set of \emph{$\rightarrow$-strongly normalising} elements is
$$\SN{\rightarrow}\eqdef\bigcap_{\mathcal T\mbox{ is \patriarchal}}{\mathcal T}
$$
\end{definition}

\begin{definition}[Bounded elements]
The set of \emph{$\rightarrow$-bounded} elements is defined as
$$\BN{\rightarrow}{}\eqdef\bigcup_{n\geq0} \BN{\rightarrow} n$$
where $\BN{\rightarrow} {n}$ is defined by induction on the
natural number $n$ as follows:
$$
\begin{array}{ll}
\BN{\rightarrow} 0&\eqdef\nf{\rightarrow}\\
\BN{\rightarrow} {n+1}&\eqdef\{M\in\mathcal A|\;\exists
n'\leq n,\suc\rightarrow M\subseteq\BN{\rightarrow} {n'}\}
\end{array}
$$
\end{definition}

\begin{lemma}\label{lem:BNSN}
If $\rightarrow$ is finitely branching, then $\BN\rightarrow{}$ is patriarchal.\\
As a consequence, $\BN{\rightarrow}{}=\SN{\rightarrow}$.
\end{lemma}

\begin{lemma}\label{lem:incl}$ $
\begin{enumerate}
\item
If $n<n'$ then $\BN\rightarrow n\subseteq\BN\rightarrow{n'}\subseteq\BN\rightarrow {}$.
In particular, $\nf\rightarrow{}\subseteq\BN\rightarrow n{}\subseteq\BN\rightarrow {}$.
\item
$\BN\rightarrow{}\subseteq\SN\rightarrow$.
\end{enumerate}
\end{lemma}

\begin{lemma}$ $\label{lem:WNSNbasic}
\begin{enumerate}
\item
$\SN\rightarrow$ is \patriarchal.
\item
If $M\in\BN{\rightarrow}{}$ then $\rightarrow
(M)\subseteq\BN{\rightarrow}{}$.\\ 
If $M\in\SN{\rightarrow}$ then
$\rightarrow (M)\subseteq\SN{\rightarrow}$.
\end{enumerate}
\end{lemma}

\begin{theorem}[Induction principle]
Given a predicate $P$ on $\mathcal A$,\\
suppose $\forall M\in\SN\rightarrow,(\forall N\in\suc\rightarrow
M,
P(N))\Rightarrow P(M)$.\\
Then $\forall M\in\SN\rightarrow,P(M)$.

When we use this theorem to prove a statement $P(M)$ for all $M$
 in $\SN\rightarrow$, we
just add $(\forall N\in\suc\rightarrow M,P(N))$ to the
assumptions, which we call the \emph{induction hypothesis}.

We say that we prove the statement \emph{by induction in
$\SN\rightarrow$}.
\end{theorem}

\begin{lemma}$ $
\label{lem:inclu}
\begin{enumerate}
\item
If $\rightarrow_1\subseteq \rightarrow_2$, then
$\nf{\rightarrow_1}\supseteq \nf{\rightarrow_2}$,
$\SN{\rightarrow_1}\supseteq \SN{\rightarrow_2}$,\\ and for all $n$,
$\BN{\rightarrow_1}{n}\supseteq \BN{\rightarrow_2}{n}$.
\item
$\nf\rightarrow=\nf{\rightarrow^+}$, $\SN{\rightarrow}= \SN{\rightarrow^+}$, and
for all $n$,
 $\BN{\rightarrow^+}{n}=\BN{\rightarrow}{n}$.
\end{enumerate}
\end{lemma}

Notice that this result enables us to use a stronger induction
principle: in order to prove $\forall M\in\SN\rightarrow,P(M)$, it now
suffices to prove 
$$\forall M\in\SN\rightarrow,(\forall
N\in\suc{\rightarrow^+}M,P(N))\Rightarrow P(M)$$
This induction
principle is called the \emph{transitive induction in
$\SN{\rightarrow}$}.

\begin{theorem}[Strong normalisation by strong simulation]
\label{Th:SNbySimul} Let $\mathcal R$ be a relation between
$\mathcal A$ and $\mathcal B$, equipped with the reduction
relations $\rightarrow_{\mathcal A}$ and $\rightarrow_{\mathcal B}$.

If $\rightarrow_{\mathcal B}$ strongly simulates
$\rightarrow_{\mathcal A}$ through $\mathcal R$, then $\mathcal
R^{-1}(\SN{\rightarrow_{\mathcal
B}})\subseteq\SN{\rightarrow_{\mathcal A}}$.
\end{theorem}

\begin{lemma}\label{lem:lexic}
Given two reduction relations $\rightarrow_1$, $\rightarrow_2$, suppose
that $\SN{\rightarrow_1}{}$ is stable under $\rightarrow_2$.  Then
$\SN{\rightarrow_1\cup\rightarrow_2}=\SN{\co{\rightarrow_1^*}{\rightarrow_2}}\cap\SN{\rightarrow_1}$.
\end{lemma}
 
\section{Strong adjournment}
\label{section:newadjourn}
\begin{definition}
Suppose $\red_{\mathcal A}$ is a reduction relation on $\mathcal A$,
$\red_{\mathcal B}$ is a reduction relation on $\mathcal B$, $\mathcal
R$ is a relation from $\mathcal A$ to $\mathcal B$.

$\red_{\mathcal B}$ \emph{simulates the reduction lengths of
$\red_{\mathcal A}$ through} $\mathcal R$ if
$$\forall k, \forall M,N\in\mathcal A,\forall P\in \mathcal B,\
M\red^k_{\mathcal A}N\wedge M\mathcal R P\ \Rightarrow\ \exists
Q\in\mathcal B,P
\red^k_{\mathcal B}Q$$
\end{definition}

\begin{lemma}\label{lem:sssl}
Suppose $\red_{\mathcal A}$ is a reduction relation on $\mathcal A$,
$\red_{\mathcal B}$ is a reduction relation on $\mathcal B$, $\mathcal
R$ is a relation from $\mathcal A$ to $\mathcal B$.

If $\red_{\mathcal B}$ strongly simulates 
$\red_{\mathcal A}$ through $\mathcal R$, then 
$\red_{\mathcal B}$ simulates the reduction lengths of
$\red_{\mathcal A}$ through $\mathcal R$.
\end{lemma}
\begin{proof}
We prove by induction on $k$ that $\forall k, \forall M,N\in\mathcal
A^2,\forall P\in \mathcal B,\ M\red^k_{\mathcal A}N\wedge M\mathcal R
P\ \Rightarrow\ \exists Q,P \red^k_{\mathcal B}Q$.
\begin{itemize}
\item
For $k=0$: take $Q\eqdef M=N$.
\item Suppose it is true for $k$ and take $M\red_{\mathcal
A}M'\red_{\mathcal A}^{k}N$.  The strong simulation gives $P'$ such
that $P\red^+_{\mathcal B}P'$ and $M'\mathcal R P'$.  The induction
hypothesis gives $Q'$ such that $P'\red_{\mathcal B}^{k}Q'$.  Then it
suffices to take the prefix $P\red_{\mathcal B}^{k+1}Q$ (of length
$k+1$) of $P\red^+_{\mathcal B}P'\red_{\mathcal B}^{k}Q'$.
\end{itemize}
\end{proof}

\begin{lemma}\label{lem:charBN}
$
\forall n,\forall M,\quad(\forall k,\forall N, M\red^k N\Rightarrow
k\leq n)\quad \Longleftrightarrow\quad M\in\BN\red n
$
\end{lemma}
\begin{proof}
By transitive induction on $n$.
\begin{itemize}
\item
For $n=0$: clearly both sides are equivalent to $M\in\nf\red$.
\item
Suppose it is true for all $i\leq n$.

Suppose $\forall k,\forall N, M\red^k N\Rightarrow
k\leq n+1$. Then take $M\red M'$ and assume $M'\red^{k'} N'$.
We have $M\red^{k'+1} N'$ so from the hypothesis we derive 
$k'+1\leq n+1$, \ie $k'\leq n$. We apply the induction hypothesis on
$M'$ and get $M'\in\BN\red n$. By definition of $\BN\red {n+1}$ we get
$M\in\BN\red {n+1}$.

Conversely, suppose $M\in\BN\red {n+1}$ and $M\red^k N$. We must prove
that $k\leq n+1$. If $k=0$ we are done. If $k=k'+1$ we have $M\red
M'\red^{k'} N$; by definition of $\BN\red {n+1}$ there is $i\leq n$
such that $M'\in\BN\red i$, and by induction hypothesis we have
$k'\leq i$; hence $k=k'+1\leq i+1\leq n+1$.
\end{itemize}
\end{proof}

\begin{theorem}\label{th:simlengbound}
Suppose $\red_{\mathcal A}$ is a reduction relation on $\mathcal A$,
$\red_{\mathcal B}$ is a reduction relation on $\mathcal B$, $\mathcal
R$ is a relation from $\mathcal A$ to $\mathcal B$.

If $\red_{\mathcal B}$ simulates the reduction lengths of
$\red_{\mathcal A}$ through $\mathcal R$, then 
$$\forall n,\mathcal R^{-1}(\BN{\red_{\mathcal B}}{n})\subseteq \BN{\red_{\mathcal
    A}}{n}\quad(\subseteq \SN{\red_{\mathcal A}})$$
\end{theorem}
\begin{proof}
Suppose $N\in\BN{\red_{\mathcal B}}{n}$ and $M\mathcal R N$.
If $M\red_{\mathcal A}^k M'$ then by simulation $N\red_{\mathcal B}^k
N'$ so by Lemma~\ref{lem:charBN} we have $k\leq n$. Hence by (the
other direction of) Lemma~\ref{lem:charBN} we have 
$M\in\BN{\red_{\mathcal A}}{n}$.
\end{proof}

\begin{definition}
Let $\rightarrow_1$ and $\rightarrow_2$ be two reduction relations on
$\mathcal A$.\\
The relation $\rightarrow_1$ can be \emph{strongly adjourned with respect to}
$\rightarrow_2$ if\\ whenever $M\rightarrow_1 N \rightarrow_2 P$ there
exists $Q$ such that $M\rightarrow_2 Q (\rightarrow_1\cup\rightarrow_2)^+ P$.
\end{definition}

\begin{theorem}\label{th:adjbound}
Let $\rightarrow_1$ and $\rightarrow_2$ be two reduction relations on
$\mathcal A$.
If $\nf{\rightarrow_2}\subseteq\nf{\rightarrow_1}$ and
$\rightarrow_1$ can be strongly adjourned with respect to $\rightarrow_2$ then
$\BN{\rightarrow_2}{}\subseteq\BN{\rightarrow_1\cup\rightarrow_2}{}$.
\end{theorem}
\begin{proof}
From Theorem~\ref{th:simlengbound}, it suffices to show that
$\rightarrow_2$ simulates the reduction lengths of $\rightarrow_1\cup\rightarrow_2$
through the identity. We show by induction on $k$ that
$$\forall k, \forall M,N,\ M(\rightarrow_1\cup\rightarrow_2)^k N
\Rightarrow\ \exists Q,M \rightarrow_2^k Q$$
\begin{itemize}
\item
For $k=0$: take $Q\eqdef M$
\item
For $k=1$: If $M\rightarrow_2 N$ take $Q\eqdef N$; if 
$M\rightarrow_1 N$ use the hypothesis
$\nf{\rightarrow_2}\subseteq\nf{\rightarrow_1}$ to produce $Q$ such
that $M\rightarrow_2 Q$.
\item Suppose it is true for $k+1$ and take
$M(\rightarrow_1\cup\rightarrow_2)P(\rightarrow_1\cup\rightarrow_2)^{k+1}
N$.

The induction hypothesis provides $T$ such that
$P\rightarrow_2^{k+1}T$, in other words $P\rightarrow_2
S\rightarrow_2^{k}T$.

If $M\rightarrow_2 P$ we are done. If $M\rightarrow_1 P$ we use the
hypothesis of adjournment to transform $M\rightarrow_1 P\rightarrow_2
S$ into $M\rightarrow_2 P'(\rightarrow_1\cup\rightarrow_2)^+ S$. Take
the prefix $P'(\rightarrow_1\cup\rightarrow_2)^{k+1}R$ (of length
$k+1$) of $P'(\rightarrow_1\cup\rightarrow_2)^+ S\rightarrow_2^{k}T$,
and apply on this prefix the induction hypothesis to get
$P'\rightarrow_2^{k+1}R$. We thus get $M\rightarrow_2^{k+2} R$.
\end{itemize}
\end{proof}

\section{Proofs}
\label{section:proofs}
\begin{ann}{Lemma}{lem:kaco}
$\co{\kaco}\gred\subseteq \co\gred{\Rewplus{\betiii}}$
\end{ann}
\begin{proof}
By induction on the reduction step $\gred$.
\begin{itemize}
\item For the base case where the $\betb$-reduction is a
$\betii$-reduction, we have\linebreak $M\kaco\ml x N P\gred \subst {N} x
P$ with $x\in\FV N$ or $P$
has no marked redex as a subterm. We do a case analysis on the reduction step $M\kaco\ml x N P$.

If $M=\ml x {N'} P\kaco\ml x N P$ because $N'\Rew{\betiii}N$ then
$\ml x {N'} P\gred \subst {N'} x P\Rew{\betiii}\subst {N} x P$.

If $M=\ml x N {P'}\kaco\ml x N P$ because
$P'\kaco P$, then it means that $P$ has a marked redex as a subterm, so we must
have $x\in\FV N$. Hence $\ml x N {P'}\gred\subst {N} x
{P'}\Rewplus{\betiii}\subst {N} x P$.
\item
For the base case where the $\betb$-reduction is a
$\beti$-reduction, we have\linebreak
$M\kaco(\l x.N)\ P\gred \subst {N} x P$ with $x\in\FV N$ or $P$
has no marked redex as a subterm. We do a case analysis
on the reduction step $M\kaco(\l x.N)\ P$.

If $M=M'\ P\kaco(\l x.N)\ P$ because $M'\kaco\l x.N$
then $M'$ must be of the form $\l x.M''$ with $M''\kaco N$.
Then $(\l x.M'')\ P\gred\subst {M''} x P$ (in case $P$ has a marked
subterm, notice that $x\in\FV{N}\subseteq\FV {M''}$), and $\subst
{M''} x P\Rew{\betiii}\subst {N} x P$.

If $M=(\l x.N)\ P'\kaco(\l x.N)\ P$ because $P'\kaco P$,
then it means that $P$ has a marked redex as a subterm, so we must
have $x\in\FV N$. Hence $(\l x.N)\ P'\gred\subst {N} x
{P'}\Rewplus{\betiii}\subst {N} x P$.
\item
The closure under $\lambda$ is straightforward.
\item For the closure under application, left-hand side, we have
$M\kaco N\ P\gred {N'}\ P$ with $N\gred {N'}$. We do a case analysis
on the reduction step $M\kaco N\ P$.

If $M=M'\ P \kaco N\ P$ with $M' \kaco N$, the induction
hypothesis gives $M'\co{\gred}{\Rewplus{\betiii}}N'$ and the weak contextual
closure gives $M'\ P\co{\gred}{\Rewplus{\betiii}}N'\ P$.

If $M=N\ P'\kaco N\ P$ with $P' \kaco P$, we can also
derive $N\ P'\gred {N'}\ P'\Rew{\betiii}N'\ P$.
\item For the closure under application, right-hand side, we have
$M\kaco N\ P\gred N\ P'$ with $P\gred P'$. We do a case analysis
on the reduction step $M\kaco N\ P$.

If $M=M'\ P \kaco N\ P$ with $M' \kaco N$, we can also
derive $M'\ P\gred M'\ P'\Rew{\betiii}N\ P'$.

If $M=N\ M'\kaco N\ P$ with $M' \kaco P$, the induction
hypothesis gives $M'\co{\gred}{\Rewplus{\betiii}}P'$ and the weak contextual
closure gives $N\ M'\co{\gred}{\Rewplus{\betiii}}N\ P'$.
\item For the closure under marked redex we have $M\kaco \ml
x P N \gred \ml x P {N'}$ with $N \gred {N'}$. We do a case analysis on
the reduction step $M\kaco\ml x P N $.

If $M=\ml x {P'} N \kaco \ml x P N$ because $P'
\Rew{\betiii} P$, we can also derive $\ml x {P'} N \gred \ml x {P'}
{N'}\Rew{\betiii} \ml x P {N'}$.

If $M=\ml x P {M'}\kaco \ml x P N$ with $M' \kaco
N$, the induction hypothesis gives $M'\gred Q\Rewplus{\betiii}N'$ and
the weak contextual closure gives $\ml x P {M'}\gred\ml x P
Q\Rewplus{\betiii}\ml x P{N'}$.
\end{itemize}
\end{proof}

\begin{ann}{Lemma}{lem:sigmgred}
$\co{\Rew{\sigmm\activ}}{\gred}\subseteq \co{\gred}{\Rewn{\sigmm\activ}}$
\end{ann}
\begin{proof}
By induction on the reduction step $\gred$.
\begin{itemize}
\item For the first base case, we have $M\Rew{\sigmm\activ}(\l x.N)\
P\gred\subst N x P$ with $x\in\FV N$ or $P$ has no marked subterm. 
Since root $\sigmm\activ$-reduction produces
neither $\l$-abstractions nor applications at the root, note that $M$
has to be of the form $(\l x.N')\ {P'}$, with either
$N'\Rew{\sigmm\activ}N$ (and $P'=P$) or $P'\Rew{\sigmm\activ}P$ (and
$N'=N$).
In both cases, $x\in\FV N\subseteq\FV {N'}$ or $P'$ has no marked
subterm, so we also have $(\l x.N')\ {P'}\gred
\subst{N'}x{P'}\Rewn{\sigmm\activ}\subst N x P$.
\item For the second base case, we have $M\Rew{\sigmm\activ}\ml x N
P\gred\subst N x P$ with $x\in\FV N$ or $P$ has no marked subterm.  We
do a case analysis on $M\Rew{\sigmm\activ}\ml x N P$.

If $M=\ml {x'}{M_1} {\ml x {M_2} P}\Rew{\sigmm}
\ml x {\ml {x'}{M_1} {M_2}} P$ with $N=\ml {x'}{M_1} {M_2}$, we also
have $M=\ml {x'}{M_1} {\ml x {M_2} P}\gred \ml {x'}{M_1} {\subst
{M_2}{x}P}=\subst N x P$.

If $M=(\l x.N)\ P\Rew{\activ}\ml x N P$ then  
$M\gred\subst N x P$.

If $M=\ml x {N'} {P'}\Rew{\sigmm\activ}\ml x N P$ with either
$N'\Rew{\sigmm\activ}N$ (and $P'=P$) or $P'\Rew{\sigmm\activ}P$ (and
$N'=N$), we have, in both cases, $x\in\FV N\subseteq\FV {N'}$ or $P'$
has no marked subterm, so we also have $(\l x.N')\ {P'}\gred
\subst{N'}x{P'}\Rewn{\sigmm\activ}\subst N x P$.
\item
The closure under $\l$ is straightforward.
\item
For the closure under application, left-hand side, we have
$Q\Rew{\sigmm\activ}M\ N\gred M'\ N$ with $M\gred M'$. We
do a case analysis on
$Q\Rew{\sigmm\activ}M\ N$.

If $Q=M''\ N\Rew{\sigmm\activ}M\ N$ with $M''\Rew{\sigmm\activ}M$, the
induction hypothesis provides $M''\co{\gred}{\Rewn{\sigmm\activ}}M'$
so $M''\ N\co{\gred}{\Rewn{\sigmm\activ}}M'\ N$.

If $Q=M\ {N'}\Rew{\sigmm\activ}M\ N$ with $N'\Rew{\sigmm\activ}N$, we
also have $M\ {N'}\gred M'\ {N'}\Rew{\sigmm\activ}M'\ N$.
\item
For the closure under application, right-hand side, we have
$Q\Rew{\sigmm\activ}M\ N\gred M\ {N'}$ with $N\gred {N'}$. We
do a case analysis on
$Q\Rew{\sigmm\activ}M\ N$.

If $Q=M'\ N\Rew{\sigmm\activ}M\ N$ with $M'\Rew{\sigmm\activ}M$, we
also have $M'\ N\gred M'\ {N'}\Rew{\sigmm\activ}M\ {N'}$.

If $Q=M\ N''\Rew{\sigmm\activ}M\ N$ with $N''\Rew{\sigmm\activ}N$, the
induction hypothesis provides $N''\co{\gred}{\Rewn{\sigmm\activ}}N'$
so $M\ N''\co{\gred}{\Rewn{\sigmm\activ}}M\ N'$.
\item For the closure under marked redex, the $\gred$-reduction can
only come from the right-hand side because of the weak contextual
closure ($\gred$ does not reduce under $\mlname$), so we have
$Q\Rew{\sigmm\activ}\ml y M P\gred\ml y M {P'}$ with $P\gred {P'}$.
We do a case analysis on $Q\Rew{\sigmm\activ}\ml y M P$.

If $Q={\ml x {M'} {\ml y N P}}\Rew{\sigmm}{\ml y {\ml x {M'} N}
P}$ with $M=\ml x {M'} N$, we also have
$Q={\ml x {M'} {\ml y N P}}\gred{\ml x {M'} {\ml y N {P'}}}
\Rew{\sigmm}{\ml y {\ml x {M'} N} {P'}}$.

If $Q=(\l y.M)\ P\Rew{\activ}\ml y M P$, then we also have 
$Q=(\l y.M)\ P\gred (\l y.M)\ {P'}\Rew{\activ}\ml y M {P'}$.
\end{itemize}
\end{proof}

\end{document}